\documentclass[12pt,letterpaper]{article}
\author{}
\usepackage{natbib,hyperref}

\usepackage[ left=1in, top=1in, right=1in, bottom=1in]{geometry}
\usepackage{graphicx,bm,colonequals,amsmath,amssymb,url,xcolor,bbm}
\usepackage{array,tabularx,multirow}
\usepackage{enumitem}
\usepackage[font={footnotesize}]{caption,subcaption}
\usepackage[utf8]{inputenc}
\usepackage{enumitem}
\usepackage{soul} 
\usepackage{placeins} 
\usepackage[normalem]{ulem}
\allowdisplaybreaks
\graphicspath{{plots/}}

\usepackage{xr}
\makeatletter
\newcommand*{\addFileDependency}[1]{
  \typeout{(#1)}
  \@addtofilelist{#1}
  \IfFileExists{#1}{}{\typeout{No file #1.}}
}
\makeatother

\usepackage{mathtools}
\mathtoolsset{showonlyrefs} 

\bibpunct[, ]{(}{)}{;}{a}{,}{,}
   
\usepackage{amsthm}
\newtheoremstyle{propstyle} 
    {2mm}                    
    {1mm}                    
    {\itshape}                   
    {}                           
    {\scshape}                   
    {.}                          
    {.5em}                       
    {}  
\theoremstyle{propstyle}

\theoremstyle{propstyle}

\theoremstyle{propstyle}

\theoremstyle{propstyle}

\theoremstyle{propstyle}

\usepackage{array,framed}
\newcounter{algorithm}

\makeatletter
\renewcommand{\paragraph}{%
  \@startsection{paragraph}{4}%
  {\z@}{2ex \@plus 1ex \@minus .2ex}{-1em}%
  {\normalfont\normalsize\bfseries}%
}
\makeatother

\DeclareMathAlphabet\mathbfcal{OMS}{cmsy}{b}{n}

\newcommand{\ba}{\mathbf{a}}

\newcommand{\bs}{\mathbf{s}}

\newcommand{\by}{\mathbf{y}}

\newcommand{\bz}{\mathbf{z}}

\newcommand{\bF}{\mathbf{F}}

\newcommand{\order}{\mathcal{O}}
\newcommand{\normal}{\mathcal{N}}

\title{Ordered conditional approximation of Potts models}

\author{Anirban Chakraborty\thanks{Department of Statistics, Texas A\&M University} \and Matthias Katzfuss\footnotemark[1] \thanks{Corresponding author: \texttt{katzfuss@gmail.com}} \and Joseph Guinness\thanks{Department of Statistics and Data Science, Cornell University}}

\date{}

\begin{document}
\maketitle
\begin{abstract}
Potts models, which can be used to analyze dependent observations on a lattice, have seen widespread application in a variety of areas, including statistical mechanics, neuroscience, and quantum computing. To address the intractability of Potts likelihoods for large spatial fields, we propose fast ordered conditional approximations that enable rapid inference for observed and hidden Potts models. Our methods can be used to directly obtain samples from the approximate joint distribution of an entire Potts field. The computational complexity of our approximation methods is linear in the number of spatial locations; in addition, some of the necessary computations are naturally parallel. We illustrate the advantages of our approach using simulated data and a satellite image.
\end{abstract}



\section{Introduction \label{sec:intro}}

Potts models can be used to describe spatial grids for which each location falls into one of a fixed number of classes. The Ising model \citep{Ising1925} can be viewed as a special case of the Potts model with two classes. Since both of these models were originally developed to describe the magnetization of atoms in crystalline solids, they involve an inverse temperature parameter $\beta$ that controls the strength of magnetization. The idea of Potts model also coincides with the autologistic models introduced by \citet{Besag1974} for analyzing categorical data on spatial grids. Potts models have been used in a variety of areas beyond statistical physics, including neuroscience \citep{Roudi2009} and quantum computing \citep{King2018}.
The hidden Potts model, in which the model for the data depends on an unobserved Potts model configuration, has also seen multiple applications, including in medical image processing \citep{doi:10.1177/1176935117711910} and satellite image analysis \citep{Moores2020}.

Potts models are computationally challenging, as model probabilities involve a normalizing constant that becomes intractable for large spatial grids. To address this intractability, \citet{2987782} introduced a pseudo-likelihood consisting of the product of full conditional distributions, which, due to a cancellation, does not require calculation of the normalizing constant. However, the asymptotic variance of the maximum-pseudo-likelihood estimator is high when the size of the spatial grid tends to infinity \citep{stoehr2017}.  
\citet{10.2307/3088837} performed an algebraic simplification of the normalizing constant in the observed Potts model using its Markov property. Although this can work efficiently for small spatial grids, the computational expense increases exponentially as the grid size increases.
Several authors have proposed MCMC algorithms and variational methods \citep[as reviewed by][]{stoehr2017} for inference on the parameter $\beta$ of the observed Potts model. Some of these methods can also be used to sample the spatial grid for given values of the parameter $\beta$.

Inference for hidden Potts models is even more involved due to a doubly intractable likelihood. Many works have been dedicated to developing approximate Bayesian computation (ABC) algorithms for speeding up the computation by assuming an approximate structure of the likelihood of a hidden Potts model that follows certain properties.  Recently, \citet{Moores2020} have introduced a parametric functional approximate Bayesian (PFAB) algorithm for inference on the temperature parameter in the hidden Potts model, which surpasses the computational efficiency of previous methods by a long margin while preserving accurate inference.

Here, we propose an ordered conditional approximation (OCA) that essentially approximates the joint density of a Potts models as a product of conditional distributions, for which we order the spatial locations by their coordinates and then condition each point on its nearest previously ordered neighbors. Our OCA is motivated by the simpler Vecchia approximation that has been highly successful for speeding up Gaussian-process inference \citep[e.g.,][]{Vecchia1988,Stein2004,Datta2016,Guinness2016a,Katzfuss2017a,Katzfuss2018}. However, unlike the Gaussian-process inference, where the conditional distributions can be calculated as closed-form multivariate Gaussian distributions, conditional distributions under the Potts model do not yield such closed-form expressions and require computing sums whose cost increases exponentially with the grid size. To overcome this computational issue, we restrict the calculation of the sums in the conditional distributions to a sub-region of the whole grid to construct the OCA. Because of its construction, our OCA can take advantage of distributed computation structures for fast calculations. Additionally, we propose a simple mechanism using our method to draw random samples from the joint distribution of a Potts field for specific values of the parameter.
We extend our approach to hidden Potts models, to allow fast evaluation of the marginal likelihood and to enable scalable inference on the hidden configuration of the spatial grids. We also propose a Gibbs sampler as an optional tool for inference in this setting.
All OCA inference scales linearly in the total number of spatial locations for fixed tuning parameters.

The remainder of this article is organized as follows. In Section \ref{sec:potts}, we describe the Potts model and introduce its OCA. In Section \ref{sec:hiddenpotts}, we extend our OCA to the hidden Potts model. In Sections \ref{sec:numericalsimulation} and \ref{sec:realdata}, we present numerical results using simulations and a satellite image, respectively. We conclude in Section \ref{sec:conclusion}.

\section{The Potts model\label{sec:potts}}

\subsection{Definition\label{sec:mrf}}

Let $\bs_1,\ldots,\bs_n$ be a set of locations on a two-dimensional rectangular spatial grid of size $n=n_1 \times n_2$. At each location $\bs_i$, we observe one of $K$ classes, $z(\bs_i) =z_i \in \{1,\ldots, K\}$. 
The Potts model assumes that the likelihood of a class label, say $k$, at a particular location increases with the number of neighboring locations on the grid falling into the same class. This likelihood is governed by an inverse temperature parameter $\beta$, which hence controls the strength of the spatial dependence in the grid.

Specifically, the joint density (i.e., the joint probability mass function) for a particular grid configuration $\bz = (z_1,\ldots,z_n)^\top$ is given by,
\begin{equation}
    \label{eq:density}
p(\bz| \beta) = \frac{\exp(-H(\bz,\beta))}{N_\beta}, 
\end{equation}
where
\begin{equation}
\label{eq:hamiltonian}
H(\bz,\beta) = \displaystyle \beta S(\bz)
\end{equation}
is the Hamiltonian function, 
\begin{equation}
    \label{eq:sum}
S(\bz)= \sum_{i\sim j} 1_{\{z_i= z_j\}}
\end{equation}
is the summary statistics, and
\begin{equation}
    \label{eq:constant}
    N_\beta = \sum_{\ba\in \{1,\ldots, K\}^n}\exp(-H(\ba, \beta))
\end{equation}
is the normalizing constant.

\begin{figure}[htp]
\includegraphics[width=\textwidth]{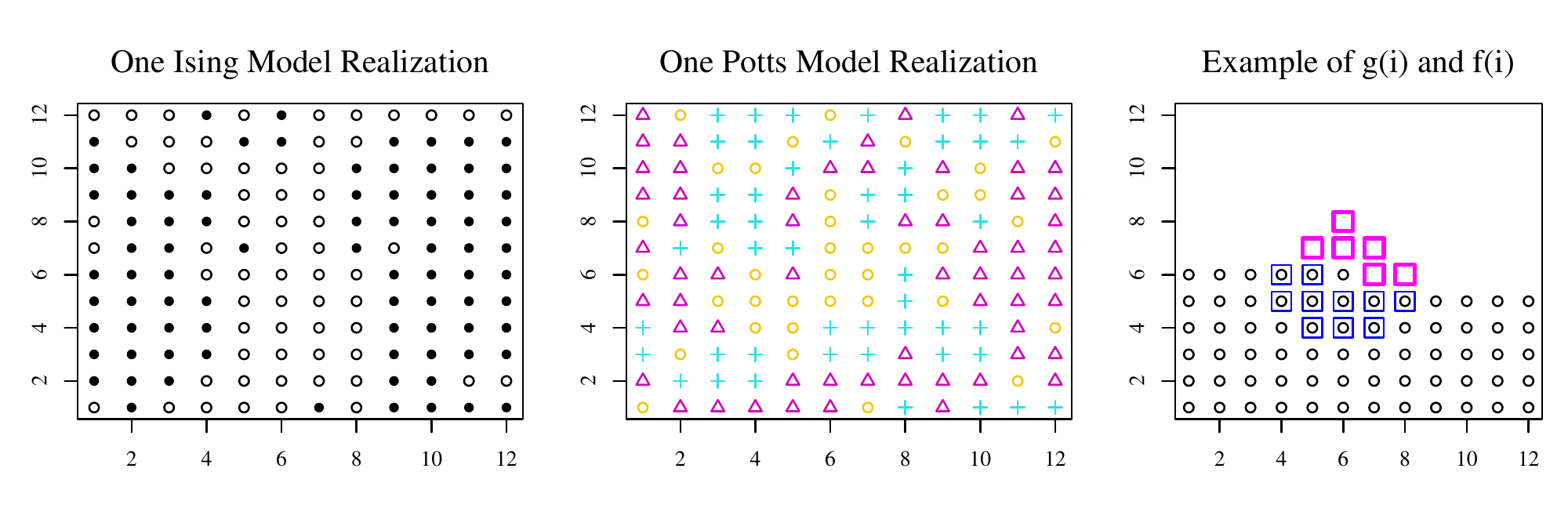}
\caption{For a spatial field of size $n=12 \times 12 = 144$, realizations of a Potts model with inverse temperature $\beta=0.35$ and $K=2$ (i.e., Ising model) at the left and $K=3$ classes in the middle panel.
Right panel: For the $i=66$th location, the points that are previous in lexicographic ordering (circles), and the $m_g=10$ nearest previously ordered points (blue boxes) and the $m_f=6$ nearest subsequently ordered points (pink boxes).}
\label{fig:realizedPotts}
\end{figure}

In \eqref{eq:sum},
$i \sim j$ means that $s_i$ and $s_j$ are neighbors. Throughout, we consider a first-order neighborhood structure consisting, for each (interior) pixel, of the four pixels immediately above, below, left, and right. More on the neighbor structure can be found at \citet{stoehr2017}.
The Ising model is a special case of the Potts model with $K=2$ states.
See Figure \ref{fig:realizedPotts} for examples of realizations of the Ising and Potts models.

For a particular grid configuration, calculation of the joint density in \eqref{eq:density} involves evaluating the normalizing constant in \eqref{eq:constant}, which requires summing over the $K^n$ possible states, and is thus computationally infeasible for even moderately large $n$.  \citet{2987782} introduced a pseudo-likelihood consisting of the product of full conditional distributions,
$\prod_{i=1}^n p(z_i | \bz_{-i},\beta )$,
which, due to a cancellation, does not require calculation of the normalizing constant. 

\subsection{Ordered conditional approximation of the Potts model\label{sec:potts_vecchia}}

To introduce our ordered conditional approximation (OCA), we first assume that the locations $\bs_1,\ldots,\bs_n$ (and hence the corresponding observations $z_1,\ldots,z_n$ in $\bz$) follow a lexicographic ordering according to their coordinates, as illustrated in the right panel of Figure \ref{fig:realizedPotts}.

Next, we obtain an ordered conditional expression of the joint density \eqref{eq:density} as a product of conditional densities:
\begin{equation}
    \label{eq:cond}
    p(\bz| \beta) = \prod_{i=1}^n p(z_i|\bz_{1:i-1}, \beta), 
\end{equation}
where the conditional distributions for the Potts model can be shown to be

\begin{equation}
    \label{eq:cond2_mod}
\textstyle {p}(z_i|\bz_{1:i-1}, \beta) =  
\frac{ 
    \displaystyle \sum_{\bz^*_{i+1:n} \in \{1, \ldots, K\}^{n-i} } 
    \exp\big( -H( (\bz_{1:i-1}, z_i, \bz^*_{i+1:n}), \beta)\big) 
}
{  
    \displaystyle \sum_{(z_i^*,\bz^*_{i+1:n}) \in \{1, \ldots, K\}^{n-i+1} } 
    \exp\big( -H ( (\bz_{1:i-1}, z_i^*, \bz^*_{i+1:n}), \beta) \big) 
}.
\end{equation}

We propose to approximate the expression in \eqref{eq:cond2_mod} by considering subvectors of $\bz_{1:i-1}$ and $\bz^*_{i+1:n}$. Define $g(i) \subset \{1,\ldots,i-1\}$ and $f(i) \subset \{i+1,\ldots,n\}$ (i.e., the blue and magenta boxes in Figure \ref{fig:realizedPotts}) of size $m_g =|g(i)|$ and $m_f = |f(i)|$, respectively. Then define the modified Hamiltonian for $V_i := \{ g(i), i, f(i) \}$ as
\begin{equation}\label{eq:mod_hamiltonian}
H_i(\bz,\beta) = \beta \sum_{\underset{j,k\in V_i}{j \sim k} } 1_{\{z_i= z_j\}}.
\end{equation}
Since $H_i$ depends only on a subset of the full set of states $\bz$, if we replace $H$ with $H_i$ in \eqref{eq:cond2_mod}, 
the sums over $\{1,\ldots,K\}^{n-i}$ and $\{1,\ldots,K\}^{n-i+1}$ can be replaced by sums over $\{1,\ldots,K\}^{m_f}$ and $\{1,\ldots,K\}^{m_f+1}$, yielding
\begin{equation}
    \label{eq:cond3_mod}
\textstyle {\hat{p}}(z_i|\bz_{1:i-1}, \beta) =  
\frac{ 
    \displaystyle \sum_{\bz^*_{f(i)} \in \{1, \ldots, K\}^{m_f} } 
    \exp\big( -H_i( (\bz_{1:i-1}, z_i, \bz^*_{i+1,n}), \beta)\big) 
}
{  
    \displaystyle \sum_{(z_i^*,\bz^*_{f(i)}) \in \{1, \ldots, K\}^{m_f+1} } 
    \exp\big( -H_i( (\bz_{1:i-1}, z_i^*, \bz^*_{i+1:n}), \beta) \big) 
},
\end{equation}
which is much less costly to evaluate, as long as $m_f$ is small. Note that while the expressions include $\bz_{i+1:n}^*$, the states that
are not part of $\bz_{f(i)}$ do not enter into $H_i$, so their values
do not matter. In addition, since $\bz_{g(i)}$ is a smaller vector than $\bz_{1:i-1}$, there are fewer terms in each individual evaluation of $H_i$, providing further computational speedups. Furthermore, since the states $\bz_{1:i-1\setminus g(i)}$ appear in both numerator and denominator in \eqref{eq:cond2_mod}, we can ignore their contribution while defining the modified Hamiltonian in \eqref{eq:mod_hamiltonian}. A detailed derivation of \eqref{eq:cond3_mod} from \eqref{eq:cond2_mod} can be found in Appendix \ref{app:pottsmodel_calc_app}. The state vector $\bz_{g(i)}$ captures the dependence of individual $z_{i}$ on its previously ordered nearest neighbors. Although the calculation time is linear in $m_g = |g(i)|$, choosing a large value of $m_g$ will not yield better results, since calculations involving the previously ordered neighbors cancel out from both the numerators and denominators in  \ref{eq:cond3_mod}. Thus, a reasonable choice of $m_g$ can be $m_g = 2 \times m_f$.

Thus, the OCA of the Potts density in \eqref{eq:cond} can be expressed as, 
\begin{equation}
    \label{eq:oca}
 \textstyle{\hat p}(\bz|\beta) = \prod_{i=1}^n \textstyle{\hat p}(z_i|\bz_{1:i-1}, \beta). 
\end{equation}
Evaluating this expression requires $\order(n(m_f+m_g+1)K^{m_f})$ time, which is linear in the grid size $n$.  Furthermore, each of the $n$ terms in \eqref{eq:oca} can be computed independently of the remaining $n - 1$ terms. Hence, we can also take advantage of parallel computation for calculating the joint log-likelihood.

Inference on $\beta$ can proceed by replacing the exact likelihood or density $p(\bz)$ by the OCA $\hat p(\bz)$, both of which implicitly depend on $\beta$. Both maximizing the (log) likelihood or Bayesian inference on $\beta$ are possible.
In contrast to the pseudo-likelihood of \citet{2987782}, 
the OCA has the virtue that it converges to the exact likelihood as $m_g$ and $m_f$ increase to $n$; we show in Section \ref{sec:paramestnum} below that the OCA can be more accurate than the pseudo-likelihood even when using very small conditioning sets.

As a brief aside, it has been shown that other orderings (e.g., the maximum-minimum-distance, or maximin ordering) can improve over lexicographic ordering in terms of the accuracy of the Vecchia approximation of Gaussian processes \citep[e.g.,][]{Guinness2016a,Katzfuss2017a}. However, we carried out exploratory numerical experiments that indicated that both maximin and reverse maximin ordering was less accurate than lexicographic ordering for the OCA of the Potts model.

\subsection{Sampling from the Potts model using OCA\label{sec:obsocasampling}}

We also provide an algorithm for joint sampling from a Potts field using OCA. Unlike many existing algorithms based on iterative techniques such as Markov chain Monte Carlo (MCMC), our approach directly draws a joint sample from the (approximate) Potts model for the whole grid (i.e., from \eqref{eq:oca}) in a single iteration. Thus, OCA does not require a large number of iterations to ensure mixing and convergence.
For each location $i$, \eqref{eq:cond3_mod} can be rewritten as,
\begin{align}
\textstyle {\hat{p}}(z_i=k|\bz_{1:i-1}, \beta) &=  
\frac{ 
    \displaystyle \sum_{\bz^*_{f(i)} \in \{1, \ldots, K\}^{m_f} } 
    \exp\big( -H_i( (\bz_{1:i-1}, z_i=k, \bz^*_{i+1,n}), \beta)\big) 
}
{  
    \displaystyle \sum_{(z_i^*,\bz^*_{f(i)}) \in \{1, \ldots, K\}^{m_f+1} } 
    \exp\big( -H_i( (\bz_{1:i-1}, z_i^*, \bz^*_{i+1:n}), \beta) \big) 
}\\
&=  
\frac{ 
    \displaystyle \sum_{\bz^*_{f(i)} \in \{1, \ldots, K\}^{m_f} } 
    \exp\big( -H_i( (\bz_{1:i-1}, z_i = k, \bz^*_{i+1,n}), \beta)\big) 
}
{  
    \displaystyle\sum_{z^*_i = 1}^K \sum_{\bz^*_{f(i)} \in \{1, \ldots, K\}^{m_f} } 
    \exp\big( -H_i( (\bz_{1:i-1}, z_i^* = k, \bz^*_{i+1:n}), \beta) \big) 
}\\
&=\frac{c_{i,k}(\bz_{1:i-1})}{\displaystyle \sum_{k=1}^K c_{i,k}(\bz_{1:i-1})},
\label{eq:samplingfrompotts}
\end{align}
for a $K$-state Potts model. 
Thus, sequentially for each $i=1,\ldots,n$, we evaluate $c_{i,k}(\bz_{1:i-1})$ for $k=1,\ldots,K$, and then sample $z_i$ from this discrete distribution. Simulation studies to examine this method are performed in Section \ref{samplingfrompottsresult}.

\section{The hidden Potts model\label{sec:hiddenpotts}}

\subsection{Definition \label{sec:latentPotts}}

\begin{figure}[htp]
\includegraphics[width=\textwidth]{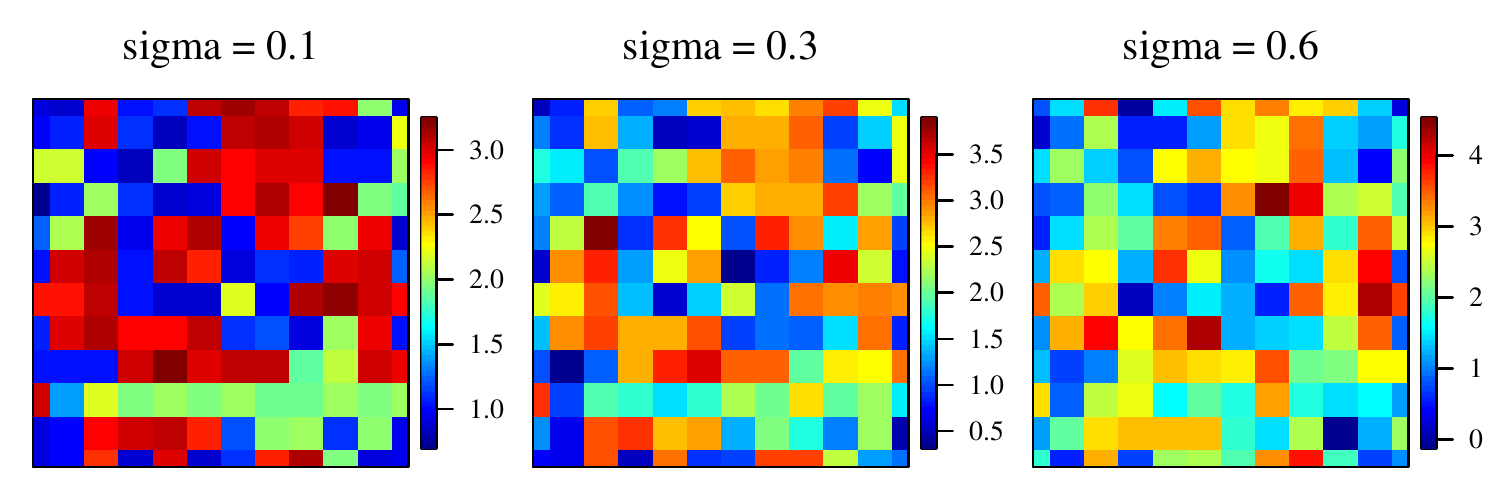}
\caption{For a single simulated hidden Potts model configuration with $K=3$ classes on a grid of size $n=12 \times 12 = 144$, noisy realizations of a hidden Potts model with a normal likelihood as in \eqref{eq:normlik} with class means $\mu_1 = 1$, $\mu_2=2$, $\mu_3=3$ and standard deviations $\sigma_1=\sigma_2=\sigma_3=\sigma$. The classes are clearly separated into red, green, and blue for low noise level $\sigma=0.1$ (left panel), but they are more difficult to distinguish for larger $\sigma=0.6$ (right).}
\label{fig:hiddenPottsplot}
\end{figure}

Let $\bz$ denote a spatial grid generated from a Potts model as in Section \ref{sec:mrf}, but now assume that $\bz$ is hidden (i.e., latent), and instead we observe a vector $\by = (y_1,\ldots, y_n)^\top$. We assume that $y_1,\ldots,y_n$ are conditionally independent given $z_1,\ldots,z_n$:
\[
p(\by|\bz) = \prod_{i=1}^n p(y_i|z_i).
\]
Various distributional assumptions for the likelihood $p(y_i|z_i)$ are possible, including a normal distribution \citep{4767596}, a Poisson distribution \citep{10.2307/3085830}, or a multivariate normal distribution for vector-valued $y_i$. For example, for the normal case, we could assume that 
\begin{equation}
    \label{eq:normlik}
y_i\,|\,\{z_i =k\} \overset{ind}{\sim} \normal(\mu_k, \sigma_k^2), \qquad k=1,\ldots,K, \quad i = 1,\ldots,n.
\end{equation}
Some realizations from such a hidden Potts model are shown in Figure \ref{fig:hiddenPottsplot}.

\subsection{OCA of the marginal likelihood\label{sec:marglik}}

Similar to Section \ref{sec:potts_vecchia}, we begin by writing the joint marginal density of the data $\by$ (i.e., the integrated likelihood) in an ordered conditional form,
\[
p(\by| \beta) = \prod_{i=1}^n p(y_i|\by_{1:i-1}, \beta),
\]
which can be written as the ratio of joint distributions
\begin{align*}
p(y_i|\by_{1:i-1},\beta) = \frac{ p(\by_{1:i}|\beta) }{ p(\by_{1:i-1}|\beta) }
&=
\frac{
\sum_{\bz^* \in \{1, \ldots, K\}^n} p(\bz^*|\beta)
\prod_{j=1}^{i} f(y_j|z^*_j)
}
{
\sum_{\bz^* \in \{1, \ldots, K\}^n} p(\bz^*|\beta)
\prod_{j=1}^{i-1} f(y_j|z^*_j)
} \\
&=
\frac{
\sum_{\bz^* \in \{1, \ldots, K\}^n} 
\exp(-H(\bz^*,\beta))
\prod_{j=1}^{i} f(y_j|z^*_j)
}
{
\sum_{\bz^* \in \{1, \ldots, K\}^n}
\exp(-H(\bz^*,\beta))
\prod_{j=1}^{i-1} f(y_j|z^*_j)
}.
\end{align*}
As before, we replace the $H$ with $H_i$, which allows us to sum
over a much smaller set,
\begin{align}
\hat p(y_i|\by_{1:i-1},\beta) 
=
\frac{
\sum\limits_{\bz^*_{g(i),i,f(i)} \in \{1, \ldots, K\}^{m_g + m_f + 1}} 
\exp(-H_i(\bz^*,\beta))
f(y_i|z^*_i)\prod\limits_{j \in g(i)} f(y_j|z^*_j)
}
{
\sum\limits_{\bz^*_{g(i),i,f(i)} \in \{1, \ldots, K\}^{m_g + m_f + 1}} 
\exp(-H_i(\bz^*,\beta))
\prod\limits_{j \in g(i)} f(y_j|z^*_j)
}.
\end{align}

Hence, by virtue of OCA, the final approximated integrated likelihood can be written as,
\begin{equation}
    \hat{p}(\by|\beta) = \prod\limits_{i=1}^n \hat{p}(y_i|\by_{g(i)}, \beta).
\end{equation}

The OCA integrated likelihood can be evaluated in $\order(n(m_g+1)(m_f+m_g+1)K^{m_f+m_g+1})$ time, which is also linear in $n$.


\subsection{OCA inference on the hidden Potts model\label{sec:samplehidden}}

We now consider the joint posterior distribution of a hidden $K$-state Potts model, which can be written as
\begin{equation}
    \label{eq:latentcond}
p(\bz | \by, \beta) = \prod_{i=1}^n p(z_i|\bz_{1:i-1},\by, \beta) = \prod_{i=1}^n p(z_i|\bz_{1:i-1},y_i,\by_{i+1:n}, \beta),
\end{equation}
since, for each location $i$, each of the previous noisy observations $\by_{1:i-1}$ are independent of 
the original states $\bz_{1:i-1}$, once $z_i$ is given.

The conditional distributions in \eqref{eq:latentcond} can be written as:
\begin{align*}
p(z_i = m\mid\bz_{1:i-1},y_i,\by_{i+1:n}) =& 
\frac{
    p(\bz_{1:i-1},z_i = m,y_i,\by_{i+1:n})
    }
    {
    \displaystyle\sum_{m \in \{1, \ldots, K\}} p(\bz_{1:i-1},m,y_i,\by_{i+1:n})
    }\\
=&\frac{
    \displaystyle\sum_{\bz^*_{i+1:n} \in \{1, \ldots, K\}^{n-i}}\prod_{j = i} ^n f(y_j \mid z^*_j) p(\bz_{1:i-1}, m, \bz^*_{i+1:n})}
    {
    \displaystyle\sum_{m \in \{1, \ldots, K\}}\sum_{\bz^*_{i+1:n} \in \{1, \ldots, K\}^{n-i}}\prod_{j = i} ^n f(y_j \mid z^*_j) p(\bz_{1:i-1}, m,\bz^*_{i+1:n})
    }.
\end{align*}

However, computation of exact posterior probability is computationally infeasible because of calculating the full Hamiltonian $H$ for all $\order(K^n)$ state configurations. 
 
To overcome this, we again replace $H$ by $H_i$, and write the approximate likelihood as: 
\begin{equation}
\label{eq:fullapproxposterior}
    \hat p(\bz|\by, \beta) = \prod_{i = 1}^n \hat{p}(z_i | \bz_{1:i-1}, y_i, \by_{f(i)}, \beta),
\end{equation}

where
\[
\hat{p}(z_i = k|\bz_{1:i-1},y_i,\by_{g(i)}, \beta) = \frac{\displaystyle\sum\limits_{\bz^*_{f(i)} \in \{1, \ldots, K\}^{m_f}}\prod f(y_j | z^*_j) \mbox{exp}(-H( (\bz_{1:i-1}, z_i=k,\bz^*_{i+1:n}),\beta))}
{\displaystyle\sum\limits_{(z_i,\bz^*_{f(i)}) \in \{1, \ldots, K\}^{m_f+1}} \prod f(y_j | z^*_j)\  \mbox{exp}(-H( (\bz_{1:i-1}, z_i,\bz^*_{1+1:n}),\beta))}.
\]
This approximate posterior probability calculation has a complexity of $\order( n(m_g+1)K^{m_f+1})$.

\subsection{Gibbs sampler for hidden Potts model\label{sec:gibbs_hidden}}

We also provide tool for Bayesian inference on hidden Potts models and parameters using the OCA that can proceed using a Gibbs sampler, for which we can sample the hidden field $\bz$ from its joint (OCA) full-conditional distribution. Let us consider the following prior distributions for the parameters of a $K$-state hidden Potts model described in Section \ref{sec:latentPotts}.
\begin{equation}\label{eq:gibbs_prior}
\mu_j \mathop{\sim}^{ind} \mathcal{N}(c_j, \sigma^2), \quad
\sigma_j^2\sim \mathcal{IG}(\alpha, \eta), \quad j = 1,\ldots,K.
\end{equation}
In addition, we assume a uniform prior on $\beta$ on the positive real line. The Gibbs sampler consists of the following steps.
\begin{enumerate}
\setlength{\itemsep}{0.01pt}
    \item Sample $\bz$, the hidden field, given $\by$, $\beta$, and the $\mu_j, \sigma_j$, as described in Section \ref{sec:samplehidden}.
    \item Update the $\mu_j, \sigma_j$ using the current hidden configuration $\bz$, by sampling from the following normal-inverse gamma posterior:
    \begin{align*}
    \sigma_j^2|\bz,\by,\beta &\sim \mathcal{IG}(\hat{\alpha}_j, \hat{\eta}_j),\\
        \mu_j|\bz, \by, \beta,\sigma_j &\sim \mathcal{N}(\hat{c}_j, \hat{\sigma}^2_j),
    \end{align*}
    where $\hat{\alpha}_j = \alpha + (n_j-1)/2$, $\hat{\eta}_j = \eta + \frac{1}{2}\sum_{z_i = j}(y_i - \bar{y_j})^2$, $\hat{c}_j= \frac{n_j\bar{y}_j/\sigma_j^2+c_j/\sigma^2}{n_j/\sigma_j^2+1/\sigma^2}$, $\hat{\sigma}^2_j =\frac{1}{n_j/\sigma_j^2+1/\sigma^2}$.
    \item Update $\beta$ using a Metropolis update based on the following un-normalized pdf:
    \[p(\beta|\bz)\propto p(\bz|\beta)p(\beta),\]
    where $p(\bz|\beta)$ is approximated as in \eqref{eq:oca}.
    \item Repeat from step 1.
\end{enumerate}

This algorithm has been illustrated with simulation studies in Section \ref{sec:hiddensim}.

\section{Numerical experiments\label{sec:numericalsimulation}}

\subsection{Simulation for observed Potts model}

\subsubsection{Parameter estimation using the OCA likelihood\label{sec:paramestnum}}

We consider the (directly observed) Potts model and its OCA described in Section \ref{sec:potts}. We demonstrate the use of the OCA likelihood in \eqref{eq:oca} for inference on the inverse temperature parameter $\beta$.

\begin{figure}[htp]
\includegraphics[width=\textwidth]{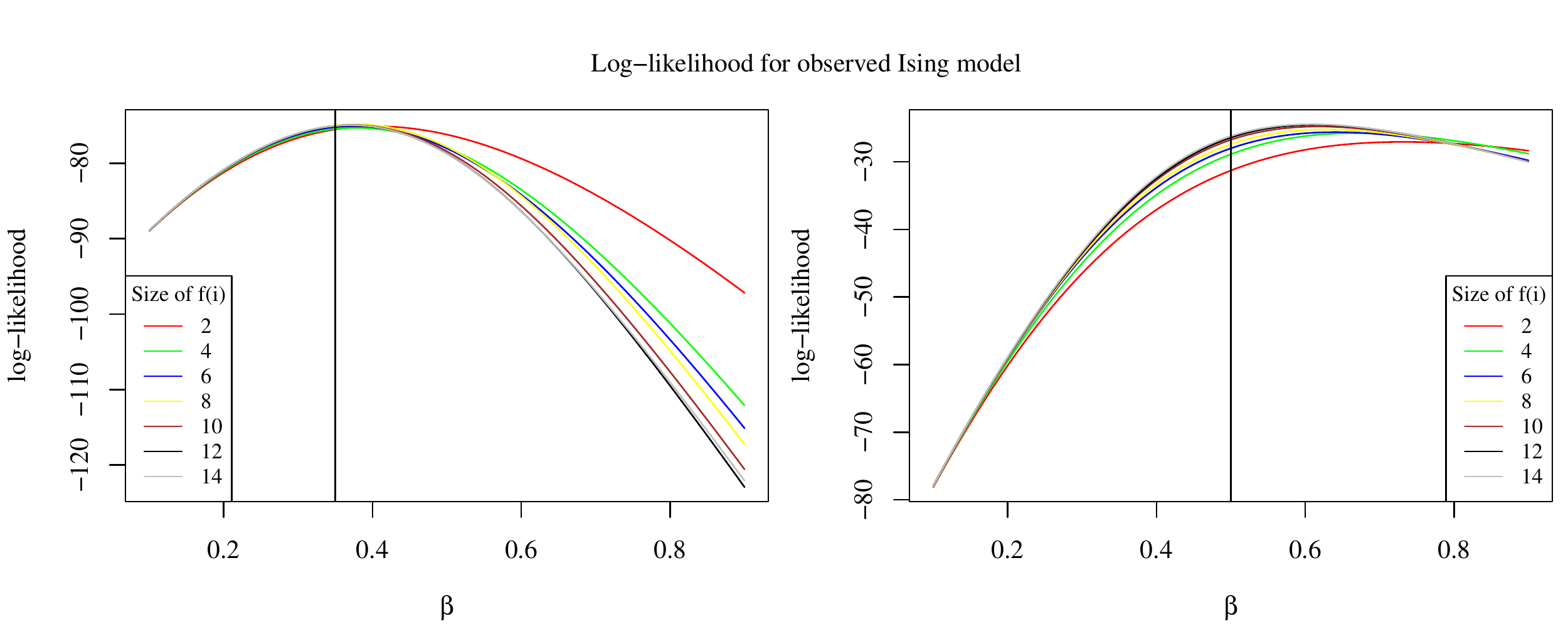}
\caption{OCAs of the likelihood $\hat{p}(\bz|\beta)$ in \eqref{eq:oca} as a function of $\beta$, for different condition set sizes $m_f$ (represented by different colours) and $m_g = 2m_f$. Data $\bz$ was simulated from an Ising model (i.e., $K=2$) on a $12\times 12$ grid with  true $\beta$ values indicated by vertical lines ($\beta = 0.35$ in the left panel; $\beta=0.5$ in the right panel).}
\label{fig:loglik-Ising}
\end{figure}

Figure \ref{fig:loglik-Ising} shows the log-likelihood for an Ising model (i.e., Potts with $K=2$) on a spatial field of size $n=12\times 12=144$ for different conditioning-set sizes $m_g$ and $m_f$. As the set size increased, the log-likelihood seemed to converge, with differences between the curves decreasing. For $m_f>4$, the OCAs log-likelihoods (and their maxima) appeared quite accurate.

\begin{figure}[htp]
\begin{center}
\begin{minipage}{0.45\textwidth}
        \centering
        \includegraphics[width=1\textwidth]{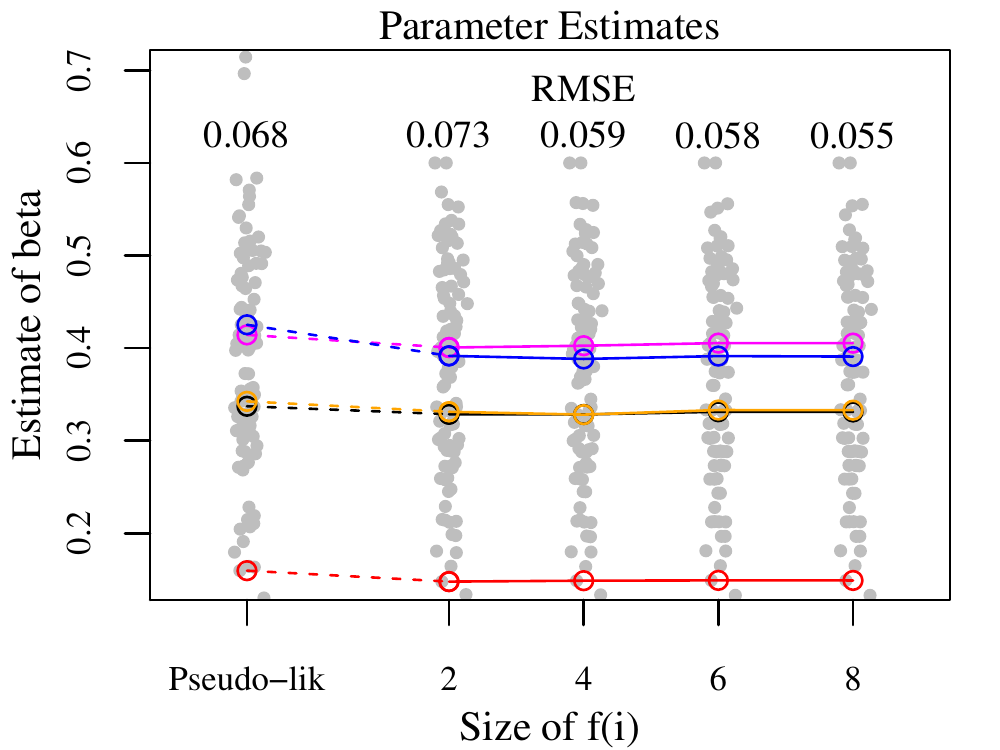}
    \end{minipage}
    \begin{minipage}{0.45\textwidth}
    \centering
        \includegraphics[width=1\textwidth]{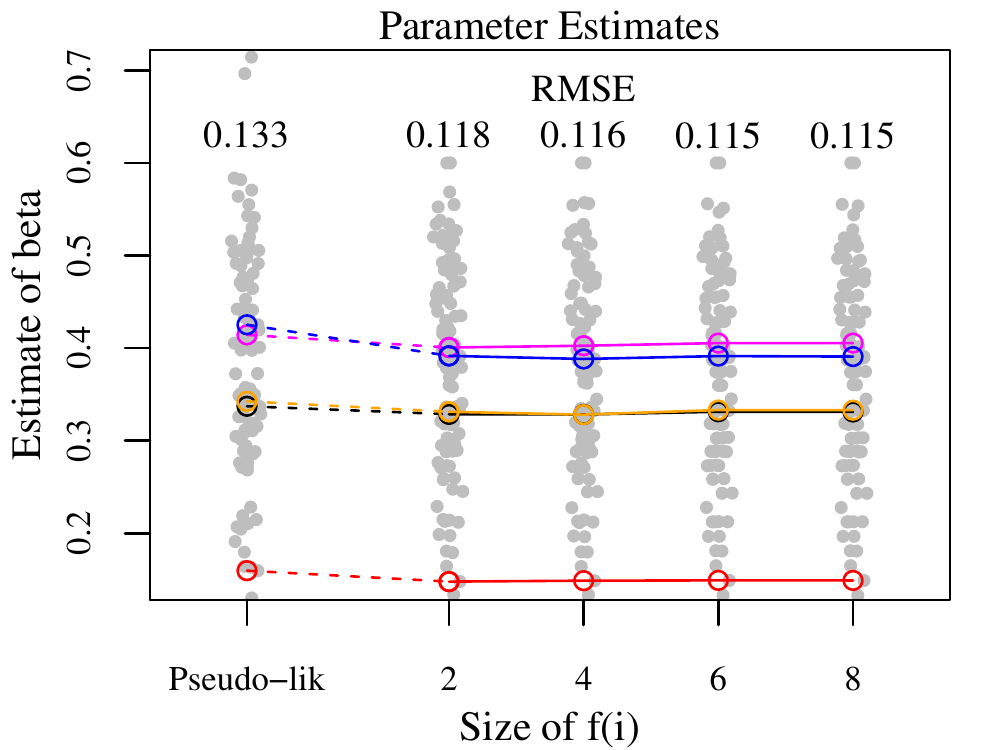}
    \end{minipage}
\caption{Parameter estimates from pseudo-likelihood and OCA for 180 simulated datasets on a $12\times 12$ grid with $\beta = 0.35$ from (a) an Ising model (left panel), and (b) a 3-state Potts model (right panel) in grey, with estimates for five datasets highlighted in color. OCA converges to the exact likelihood as size of $f(i)$ increases, and so OCA parameter estimates converge to maximum likelihood estimates.}
\label{fig:Ising}
\end{center}
\end{figure}

For a more systematic comparison, we fixed the true value of $\beta$ at 0.35 and simulated 180 datasets from the Ising model. After that, we obtained estimates of $\beta$ for each dataset by maximizing both the OCA likelihood and the pseudolikelihood described in \citet{Besag1974}. We then compared the quality of these estimates using the root mean squared error (RMSE). 
As shown in Figure \ref{fig:Ising}(a), the OCA estimates converged quickly as $m_f$ increased,  resulting in a smaller RMSE than for the pseudo-likelihood. The OCA likelihood evaluations were computationally feasible as well; even with $m_f=10$, OCA Ising likelihood evaluations were obtained in less than 60 seconds on a laptop.
 
We repeated the same set of simulations in Figure \ref{fig:Ising}(b), but instead of using the Ising model, we simulated datasets from a 3-state Potts model using the \texttt{PottsUtils} \citep{PottsUtilsR} package in \texttt{R}. OCA yielded better RMSEs than the pseudolikelihood method.

\subsubsection{Sampling from observed Potts model
\label{samplingfrompottsresult}}

We simulated 60 samples from a 3-state observed Potts model for different values of $\beta$ using the OCA likelihood on a 50$\times$50 spatial grid. The required steps for sampling have been described in Section \ref{sec:obsocasampling}. We also considered several sizes $m_f$ of the future condition sets, while setting $m_g = 2 m_f$.

\begin{figure}[htp]
\includegraphics[width=\textwidth]{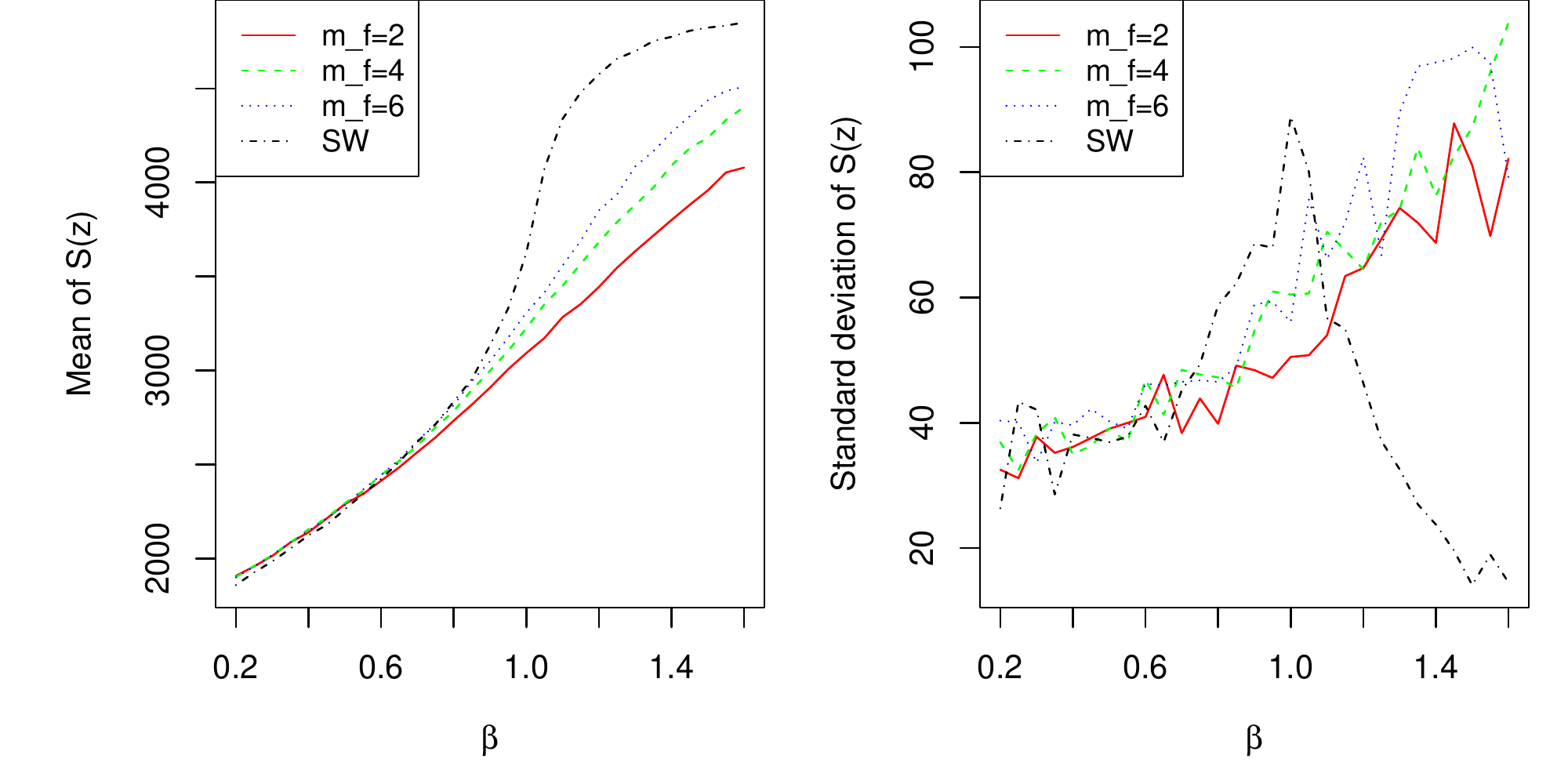}
\caption{Mean and standard deviation of summary statistics S($\bz$) for 60 observations simulated from a 3-state Potts models on a 50$\times$50 spatial grid. $m_f$ denotes different future conditional set sizes, and SW denotes the Swendsen-Wang algorithm.}
\label{fig:obsocasamplingcompare}
\end{figure}

We considered the Swendsen-Wang algorithm \citet{swendsenwang} for comparison purposes. Since the model configurations are nominal variables that indicate only the class labels and can not be compared individually, the summary statistics S($\bz$) has been computed. Our simulation results are presented in Figure \ref{fig:obsocasamplingcompare}. 

From the plot, it can be concluded that while our method performed well below the critical temperature ($\beta \leq \mbox{log}(1+\sqrt{\mbox{K}}) \approx 1$), the estimates of both mean and standard deviation for higher values of $\beta$ can differ from the Swendsen-Wang algorithm more substantially. Yet because of its simple construction, OCA likelihood can be a useful tool when sampling from small values of $\beta$. Fixing different values of the future condition set size $m_f$ = 2, 4, and 6, a single draw of random sample from a 3-state Potts model on a $50 \times 50$ spatial grid on an average required 0.01, 0.1, and 2.4 seconds, respectively.

\subsection{Simulation for hidden Potts model\label{sec:hiddensim}}
We followed our assumptions in Section \ref{sec:latentPotts} to generate realizations from a 3-state hidden Potts model on a $12\times 12$ spatial field. We assumed $\mu_j =j$ for all $j = 1,\ldots, K = 3$. Generally speaking, for simulating data from a hidden Potts model, we first simulated the latent Potts field using the \texttt{PottsUtils} \citep{PottsUtilsR} package, and then we simulated the noisy observed data conditional on the Potts field class at each pixel using the \texttt{rnorm} function in \texttt{R}.

\subsubsection{Integrated likelihood}

\begin{figure}[htp]
\includegraphics[width=\textwidth]{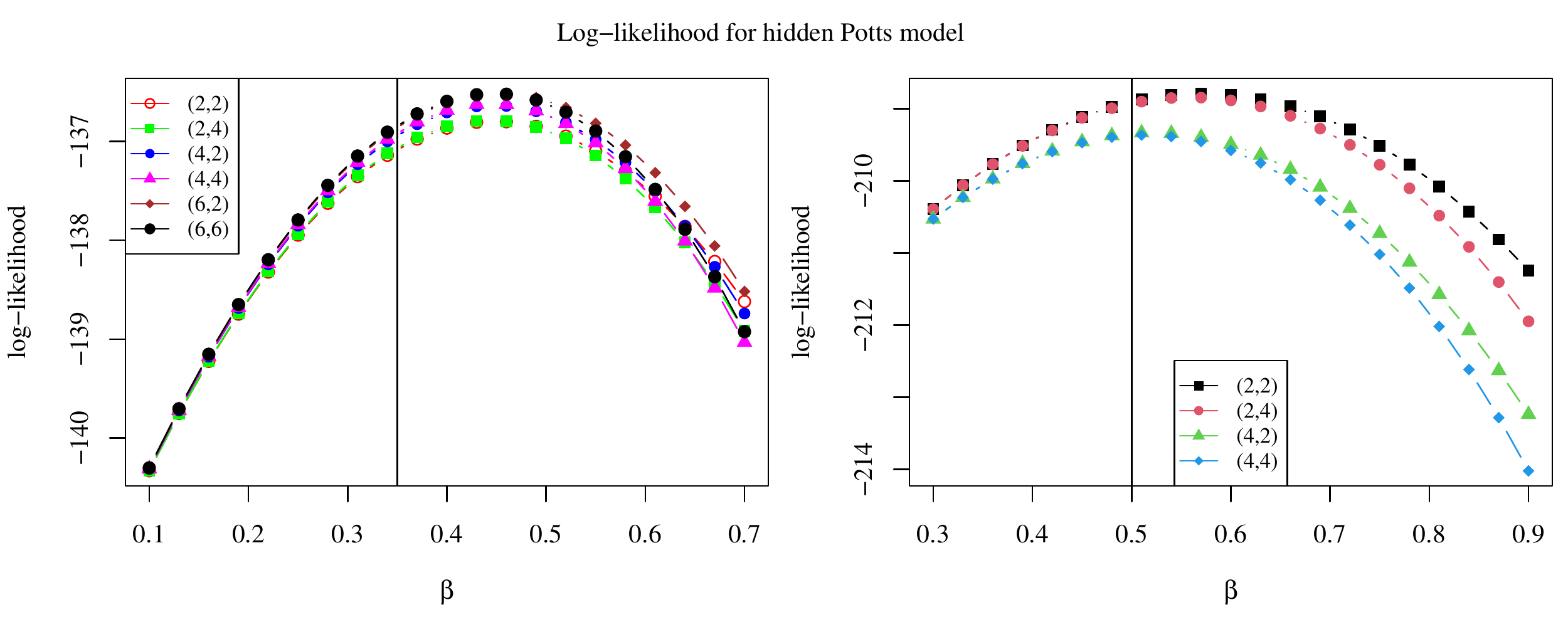}
\caption{Log-likelihood of a 3-state (at the left) and a 5-state (at the right) hidden Potts Model on a $12 \times 12$ grid with $\beta = 0.35, 0.5$, respectively. Log-likelihoods for two different condition set sizes have been plotted through two different colours as well as points. Different colors correspond to different combinations of $(m_g,m_f)$.}
\label{fig:latentPottsloglik}
\end{figure}

Figure \ref{fig:latentPottsloglik} shows integrated log-likelihoods for hidden Potts models. For data generated from a 3-state and a 5-state Potts model for two different $\beta$ values (indicated by vertical lines) with noise level $\sigma =0.25$, we calculated the OCA likelihood of the data at different values of $\beta$. The computational expense for calculating the log-likelihood grows exponentially with the size of the conditioning sets. On average, a single evaluation of the integrated likelihood takes 2 seconds with $m_f = 2$, 4 minutes with $m_f = 4$, and 10 minutes with $m_f = 6$ for a 3-state hidden Potts model.

\begin{figure}[htp]
\includegraphics[width=\textwidth]{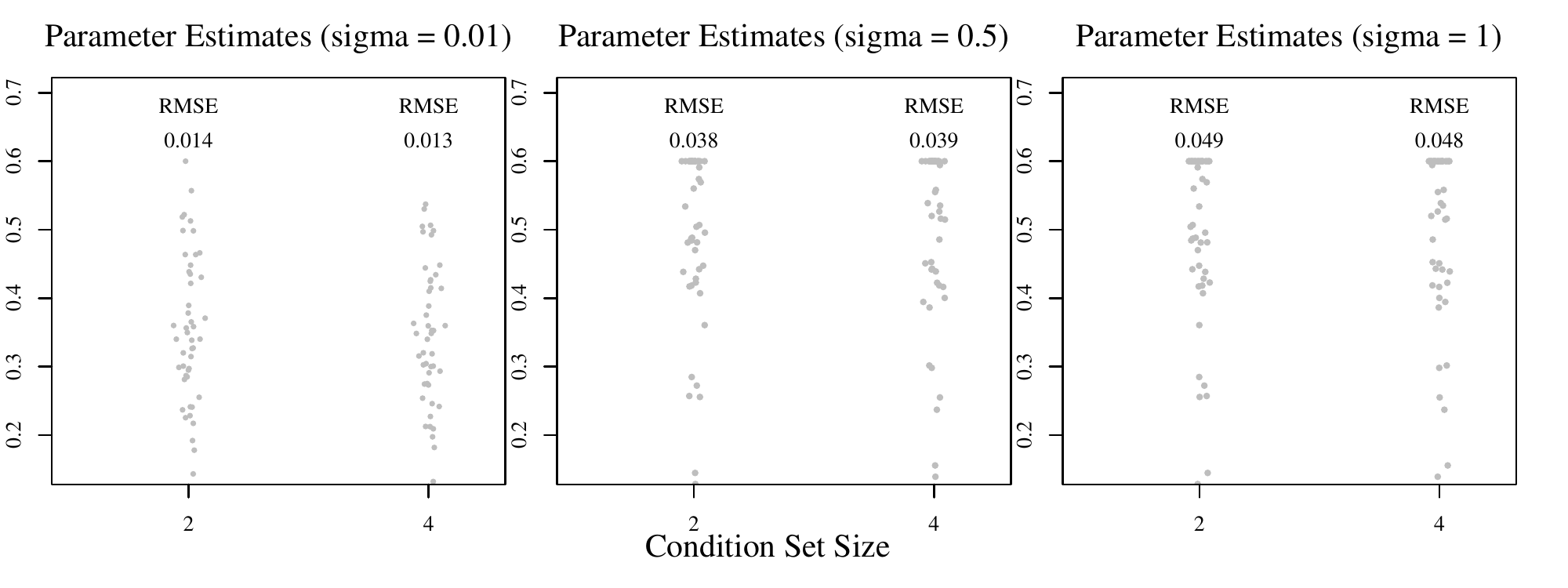}
\caption{OCA estimates for $\beta$ and corresponding RMSEs for a hidden 3-state Potts model with varying noise standard deviations. Here, $m_g = \mbox{2} m_f$.}
\label{fig:latentPottsbetas}
\end{figure}

As expected, the error in estimating $\beta$ increased with the noise level in Figure \ref{fig:latentPottsbetas}. While the RMSE was approximately 0.014 for estimating $\beta$ on a hidden 3-state Potts model with low noise, it increased multiple folds when more noise was added.  
For $m_f=m_g=2$, we obtained an estimate of $\beta$ for the hidden Potts model within 2 minutes on a laptop. Computation of the likelihood as well as optimization become expensive as $m_g$ increases. 

\begin{figure}[htp]
\begin{center}
\includegraphics[width=\textwidth]{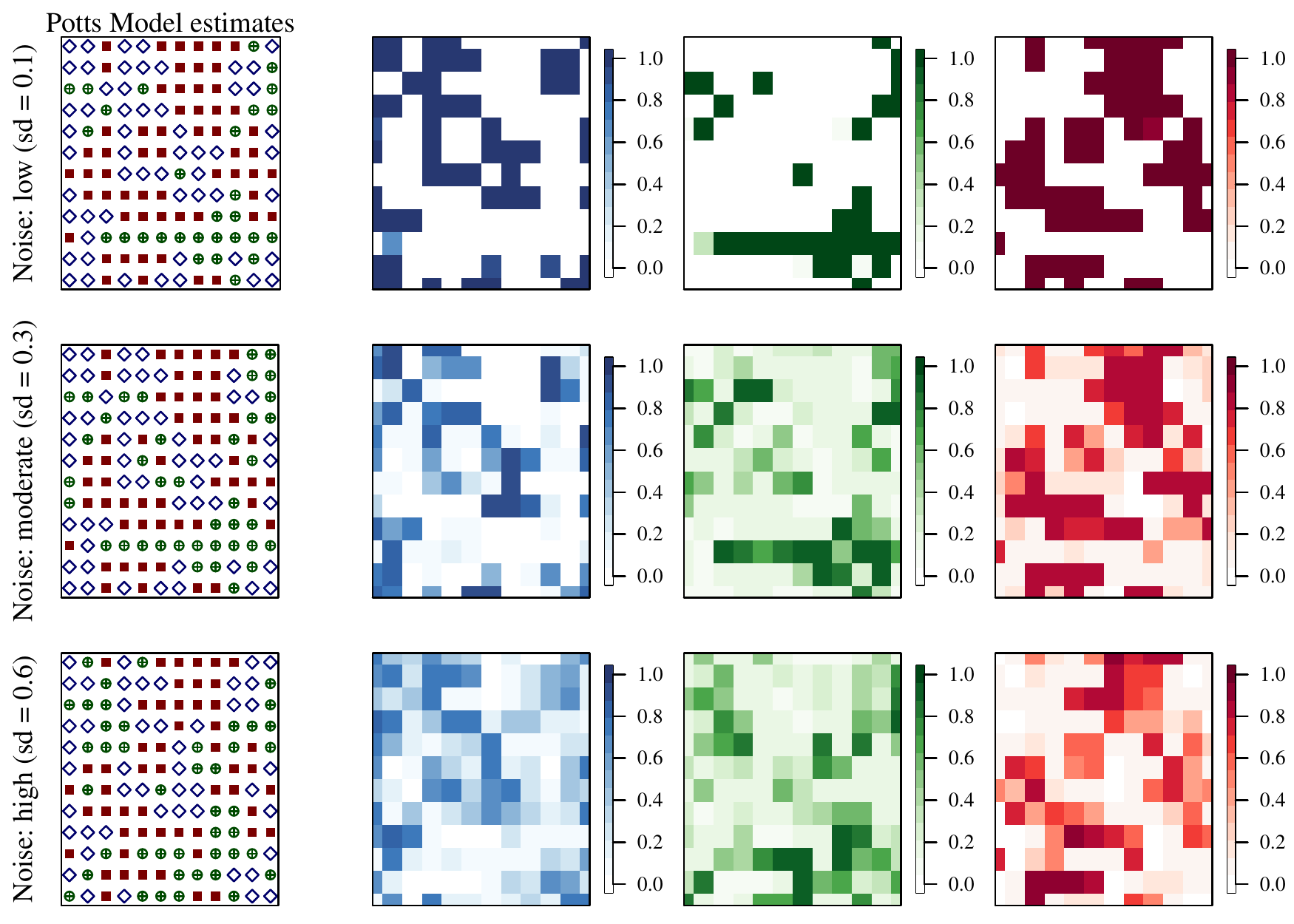}
\caption{Inference on the latent field for a 3-state hidden Potts model with three different noise levels (corresponding to the three rows). The observed data are shown in the left panel of Figure \ref{fig:hiddenPottsplot}. The first column shows the inferred highest-posterior-probability (HPP) map, while the remaining three columns show the posterior probabilities for classes 1, 2, 3, respectively.}
\label{fig:higherpostprobs}
\end{center}
\end{figure}

\subsubsection{Gibbs sampler for the hidden field\label{sec:hiddenPottsGibbs}}

We considered the Gibbs sampler described in Section \ref{sec:gibbs_hidden} for a 3-state hidden Potts model (i.e., $K = 3$). The data were generated with different noise levels, which can be found in Figure \ref{fig:hiddenPottsplot}.
For parameters of the prior distributions assumed in \eqref{eq:gibbs_prior}, we considered $\sigma = 0.1$, $\alpha  = 1.5$, $\eta = 0.135$, and $c_j = j$ for $j \in \{1,2,3\}$. 

We ran the Gibbs sampler for 8,000 iterations, with the first half considered burn-in. Our numerical results are plotted in Figure \ref{fig:higherpostprobs}, including the highest-posterior-probability (HPP) map, which for each pixel shows the class that had the highest posterior probability. The results show that, even with moderately large noise in data, our algorithm can generally extract the hidden spatial structure.

\subsubsection{Comparison with other models\label{sec:brierscorecomp}}
We used the Brier score as a tool of comparison between our method and the finite mixture model of Gaussian distributions (a brief overview of this Gaussian mixture model can be found in Appendix \ref{app:gmm_theory_app}), and the PFAB algorithm illustrated in \citet{Moores2020}. 
The Brier score measures the accuracy of probabilistic prediction for a set of mutually exclusive discrete outcomes. It is defined by,
\[BS = \frac{1}{N}\sum_{t=1}^n\sum_{j=1}^K (f_{j,t}-o_{j,t})^2,\]
where $f_{j,t}$ is probability of forecast in $j^{th}$ class for $t^{th}$ observation, and $o_{j,t}$ is an indicator function which attains the values of 1 if the $t^{th}$ observation truly belongs to class j, and 0 otherwise. Hence, by its construction, Brier score takes on small positive values when there is little mismatch between the forecast and the test data (i.e., when the calibration is good).

Using the results found in Section \ref{sec:hiddenPottsGibbs}, we calculated the Brier score for the three different methods. The result is presented in Table \ref{tab:brierscore}.

\begin{table}[htp]
    \centering
    \begin{tabular}{|c|c|c|c|}\hline
     \ & $\sigma_j$ = 0.1 & $\sigma_j$ = 0.3 & $\sigma_j$ = 0.6 \\ \hline
    GMM & 0.705 & 0.687 & 0.684\\ \hline
    OCA & 0 & 0.156 & 0.560\\ \hline 
    PFAB & 0 & 0.075 & 0.328\\ \hline 
    \end{tabular}
    \caption{Brier score for three algorithms - Gaussian mixture model (GMM), Ordered conditional approximation (OCA) and Parametric functional approximate Bayesian (PFAB) in three different signal to noise ratio (SNR)'s. $\sigma_j$ denotes the noise for the j$^{th}$ pixel. In our simulation, we have considered same noise level for all the pixels.}
    \label{tab:brierscore}
\end{table}

From our simulation results, we can interpret that in all of the cases, our method (OCA) worked well compared to the Gaussian mixture model (GMM). However, it performs poorly with respect to \citet{Moores2020} (PFAB) in terms of calibration. However, the PFAB algorithm requires $\beta$ to be a scalar random variable. If $\beta$ is vector-valued (e.g., as in the modified Potts model of \citealp{doi:10.1177/1176935117711910}), the current PFAB algorithm does not work, whereas our method could be extended to such a scenario.

\section{Analysis of a satellite image\label{sec:realdata}}

\subsection{Image classification}\label{sec:imageclassification}

\begin{figure}[htp]
\includegraphics[width=\textwidth]{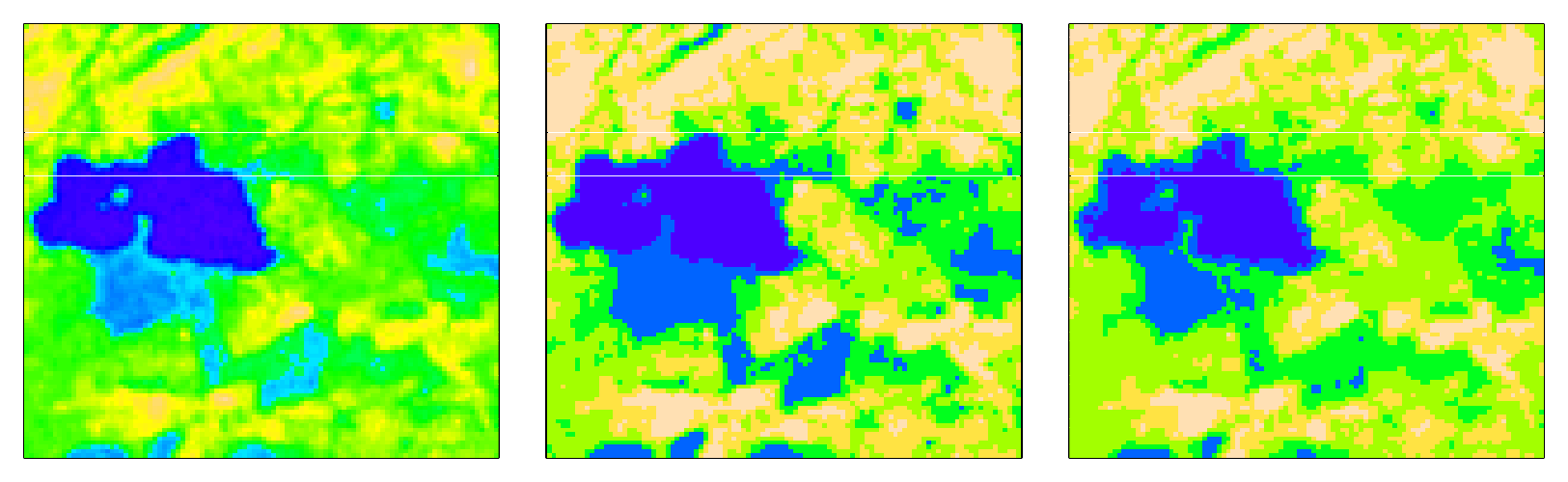}
\caption{The leftmost panel displays the original (noisy) Menteith data, the middle panel displays the highest posterior probability configuration of the latent Potts field (based on 50 Gibbs iterations), and the rightmost panel displays a single sample from the latent field. 
}
\label{fig:realdatapotts}
\end{figure}

We applied our method to the \texttt{Menteith} data in \citet{bayessR}. This is a 100$\times$ 100 satellite image of the lake of Menteith, near Stirling, Scotland. Following the description by \citet{bayessR}, our objective is to use the noisy satellite data of the lake and the land bodies surrounding it, to classify each pixel into one of six different classes of surface types, using the Gibbs sampler described in Section \ref{sec:gibbs_hidden}. We have performed a $k$-means algorithm with six classes on the dataset and used the corresponding class means and standard deviations to find the optimal value of $\beta$ based on the log-likelihood described in Section \ref{sec:hiddenpotts}. The corresponding value was then used as the initial value of $\beta$ for a Gibbs sampler as described in Section \ref{sec:samplehidden}. To mitigate the computational expense, we considered $m_f = 2$ in Section \ref{sec:potts_vecchia} and Section \ref{sec:samplehidden} while calculating the log-likelihood of $\beta$ and sampling from the posterior distributions. The original satellite picture and the HPP image are provided in Figure \ref{fig:realdatapotts}, along with one image sampled from the posterior to give a sense of the posterior uncertainty for the latent field. It is interesting to note that, while our method is able to distinguish between different classes, all the pixels are classified into a single class when we use the a Gaussian mixture model (see Appendix \ref{app:gmm_theory_app}) with six classes or components.

On a personal laptop with an Intel i5 7th gen CPU and 2 cores, a single draw of the random sample from hidden Potts model takes 2 seconds, while computation of log-likelihood of $\beta$ for the 100$\times$100 hidden Potts model takes around 5 minutes, resulting in slow Metropolis filter for sampling of $\beta$ in Section \ref{sec:hiddenPottsGibbs}. Since our log-likelihood calculation is able to take advantage of the distributed computation, we believe that this calculation can be done faster using a system with higher number of processing cores.

\subsection{Prediction for held-out pixels\label{sec:missingobservations}}

Using the same setup as in Section \ref{sec:imageclassification}, we effectively held out 10$\%$ randomly chosen pixels of the 100$\times$100 spatial grid (i.e., 1,000 pixels) by assuming each of them to have a very large (known) standard deviation of 100, which means that the information in these 1,000 data points was down-weighted and largely ignored. We then obtained the posterior distribution of the entire latent field as in step 1 in Section \ref{sec:gibbs_hidden}, in effect generating out-of-sample predictions for the ``held-out'' 1,000 pixels.

To overcome the computational limitation on a personal laptop, we considered only 100 MCMC iterations. After performing steps 1 and 2  of Section \ref{sec:gibbs_hidden} in each iteration, we consider the previously chosen pixels, and draw 100 normal samples using the predicted means and standard deviations of the corresponding classification labels for each of those pixels that we get from step 2. For example, if one of the 1000 pixels has a prediction as label 3, then for that particular pixel we draw 100 normal samples using $\mu_3$ and $\sigma_3$ that we get from Step 2 of the Gibbs sampler in Section \ref{sec:gibbs_hidden}. Then, we consider continuous rank probability score (CRPS) to measure the difference between the originally observed value of the pixels and the 100 samples. Given a probability distribution $\bF$ and a sample $y$, the CRPS is given by,
\[
   CRPS(y,\bF) = \displaystyle\int_{-\infty}^{\infty} (\bF(z) - \mathbf{1}\{y\leq z\})^2dz.
\]
We repeat this process of randomly selecting pixels and perform following steps 10 times, and take average CRPS of the whole process.
We do the same experiment using the Gaussian mixture model (GMM). 

From our experiment, the Gibbs sampler method using the OCA approach yields an average CRPS of 5.43, while the GMM yielded a much higher average CRPS of 20.36. From the definition of CRPS, it is clear that low CRPS implies that the absolute difference between the original distribution and the empirical distribution of the samples are closely situated, which means better prediction. Using this fact, we can conclude in this section that, when less information about the pixels are present (which also means high standard deviation), OCA performs better than GMM in terms of prediction.

\section{Conclusions \label{sec:conclusion}}

We introduced an ordered conditional approximation (OCA) of the likelihood of both the Potts model and hidden Potts model. OCA does not require calculation of the intractable normalising constant. We have also devised a simple algorithm to sample from the observed Potts model using OCA and numerically shown that it can be used to sample from the Potts model for a range of values of the parameter $\beta$. Our method could straightforwardly be extended to higher spatial dimensions (i.e., more than 2 coordinates) because of its simple structure of calculating multiple conditional likelihoods (refer to Section \ref{sec:potts_vecchia}). A Gibbs sampler for inference on the spatial configuration and the parameter $\beta$ has also been suggested. We have tested these methods on both simulated data and real observations.

Inference using our methods requires linear time in $n$, total number of locations in the spatial grid. We have used root mean squared error (RMSE) to show superiority of our method over pseudo-likelihood. Additionally, we have used the Brier score to show better consistency of our method compared to a Gaussian mixture model. As calculation of the likelihood slows down with increasing conditioning-set size, inference becomes infeasible for large datasets in personal laptops. As a side-note, OCA should work well if applied to the algorithm developed by \citet{Moores2020}; however, because of its redundancy, we have not performed this step.

One of the possible future directions of this work might be developing theoretical properties of OCA based on varying conditional set size. \citet{Schafer2020} have provided some theoretical bounds of conditional likelihood approximations in the context of Gaussian processes, but, to the best of our knowledge, the same results can not be applied directly in the context of Potts model. Another direction is to extend our approach to the modified Potts models, where the parameters vary based on the class labels \citep{doi:10.1177/1176935117711910}. As the latter involves more than one parameter, gradient-based optimization may be implemented to obtain the maximum likelihood estimators.

\footnotesize
\appendix
\section*{Acknowledgments}

MK and JG were partially supported by National Science Foundation (NSF) Grant DMS--1953005. MK was also partially supported by NSF Grants DMS--1654083 and CCF--1934904.


\section{Calculation of Potts model likelihood\label{app:pottsmodel_calc_app}}
Here our objective is to approximate the conditional likelihood described in \eqref{eq:cond2_mod}. It is to be noted that, for each fixed $i$, calculation of the conditional likelihood takes approximately $\order{(K^{n-i})}$ unit of time, which is infeasible for small values of $i$. For this reason, we truncate the calculation of our conditional likelihoods up to a subset of future indices for every $i$. So, we define, $f(i) \subset \{i+1,\ldots,n\}$ (i.e., the magenta boxes in Figure \ref{fig:realizedPotts}) of size $m_f = |f(i)|$. Then, we define the first modified Hamiltonian as,
\begin{equation}
    \label{eq:mod_hamil_1_app}
    \displaystyle H^*_i(\bz, \beta) = \beta \sum_{\underset{j, k \in \{1, \ldots, i, f(i)\}}{j \sim k}}\mbox{1}_{z_j = z_k}.
\end{equation}
Here we consider the first-order nearest neighborhood structure; that is, $\mbox{1}_{z_j = z_k}$ will take the value 1 only when $j \in \{ k - n_2, k - 1, k + 1, k + n_2\}$. A closer look in Figure \ref{fig:realizedPotts} suggests that the modified Hamiltonian in \eqref{eq:mod_hamil_1_app} can be written as,
\begin{equation}
    \displaystyle H^*_i(\bz, \beta) = H^\prime_i(\bz, \beta) + H_i(\bz, \beta),
\end{equation}
where $H^\prime_i(\bz, \beta) = \displaystyle \sum_{\underset {j, k \in \{1, \ldots, i -1\}\setminus g(i)}{j \sim k}}\mbox{1}_{z_j = z_k} + \sum_{\underset {k \in g(i), j\in \{1, \ldots, i -1\}\setminus g(i)}{j \sim k}}\mbox{1}_{z_j = z_k}$, and $H_i(\bz, \beta) = \displaystyle \sum_{\underset{j, k \in V_i = \{g(i), i, f(i)\}}{j \sim k}}\mbox{1}_{z_j = z_k}$ for suitable choice of $g(i) \subset \{1, \ldots, i-1\}$ with $m_g = |g(i)|$, since the locations with indices in f(i) will only have nearest neighbors from the locations with indices in $g(i)$ (the blue boxes in Figure \ref{fig:realizedPotts}). Finally we can approximate the conditional likelihood as,

\begin{align}
\label{eq:cond3_mod_app}
\textstyle {\hat{p}}(z_i|\bz_{1:i-1}, \beta) &=  
\frac{ 
    \displaystyle \sum_{\bz^*_{f(i)} \in \{1, \ldots, K\}^{m_f} } 
    \exp\big( -H^*_i( (\bz_{1:i-1}, z_i, \bz^*_{i+1,n}), \beta)\big) 
}
{  
    \displaystyle \sum_{(z_i^*,\bz^*_{f(i)}) \in \{1, \ldots, K\}^{m_f+1} } 
    \exp\big( -H*_i( (\bz_{1:i-1}, z_i^*, \bz^*_{i+1:n}), \beta) \big) 
}\\
&=\frac{ 
    \displaystyle \sum_{\bz^*_{f(i)} \in \{1, \ldots, K\}^{m_f} } 
    \exp\big( -(H^\prime_i((\bz_{1:i-1}, z_i^*, \bz^*_{i+1:n}), \beta) + H_i((\bz_{1:i-1}, z_i^*, \bz^*_{i+1:n}), \beta)), \beta)\big) 
}
{  
    \displaystyle \sum_{(z_i^*,\bz^*_{f(i)}) \in \{1, \ldots, K\}^{m_f+1} } 
    \exp\big( -(H^\prime_i((\bz_{1:i-1}, z_i^*, \bz^*_{i+1:n}), \beta) + H_i((\bz_{1:i-1}, z_i^*, \bz^*_{i+1:n}), \beta)) \big) 
}\\
&=\frac{ 
    \displaystyle \sum_{\bz^*_{f(i)} \in \{1, \ldots, K\}^{m_f} } 
    \exp\big( -H_i((\bz_{1:i-1}, z_i^*, \bz^*_{i+1:n}), \beta)\big) 
}
{  
    \displaystyle \sum_{(z_i^*,\bz^*_{f(i)}) \in \{1, \ldots, K\}^{m_f+1} } 
    \exp\big( -H_i((\bz_{1:i-1}, z_i^*, \bz^*_{i+1:n}), \beta) \big) 
},
\end{align}
since the first part of the modified Hamiltonian $H^\prime_i$ is constant in both numerator and denominator. The final expression in \eqref{eq:cond3_mod_app} yields the approximation stated in \eqref{eq:cond3_mod}.

\section{Gaussian mixture model (GMM)}\label{app:gmm_theory_app}
Suppose we have $n$ observations $\by =(y_1, y_2, \ldots, y_n)$, and the corresponding state vector $\bz = (z_1, z_2,\ldots, z_n)$ that denotes individual class labels. We assume that the observations follow the following distribution.
\begin{equation}
    \label{eq:normlik_app}
y_i\,|\,\{z_i =k\} \overset{ind}{\sim} \normal(\mu_k, \sigma_k^2), \qquad k=1,\ldots,K, \quad i = 1,\ldots,n,
\end{equation}
where $K$ is the maximum number of unique states.

The goal is to make inference about the class labels and the corresponding parameters. However, we do not have the opportunity to directly observe the class labels. Instead we only observe $\by$. For this reason, we consider some priors on the class labels and the parameters using a framework called the Gaussian mixture model (GMM). 

First, we consider the following priors for the individual means and standard deviations:
\begin{equation}\label{eq:gibbs_meansd_prior_app}
\mu_j \mathop{\sim}^{ind} \mathcal{N}(c_j, \sigma^2), \quad
\sigma_j^2\sim \mathcal{IG}(\alpha, \eta), \quad j = 1,\ldots,K.
\end{equation}

The above distributions have followed the same prior structure with the Bayesian framework illustrated in Section \ref{sec:gibbs_hidden}. However, here we do not assume that $\bz \sim \mbox{Potts}(\beta)$. Instead, we assume that,
\begin{equation}
    \label{eq:gibbs_z_prior_app}
    z_i|\pi_1,\ldots \pi_K \mathop{\sim}^{iid} multinominal(\pi_1, \ldots, \pi_K), \quad i = 1, \ldots, n.
\end{equation}
where $\Pi = \{\pi_1,\ldots \pi_K\}$ denote the individual class probabilities. We assume that the class probabilities are not known either (since we do not have any information about the class labels, as discussed earlier), and so we consider a Dirichlet prior on $\Pi$. We assume,
\begin{equation}
    \label{eq:gibbs_pi_prior_app}
    \Pi \sim Dirichlet(\alpha_1, \ldots, \alpha_K),
\end{equation}
where $\alpha_1, \ldots, \alpha_K$ are the tuning parameters that control the mean and variances of corresponding parameters. We have fixed $\alpha = \frac{1}{}K$ as the tuning parameter in the numerical simulations. For example, $\alpha = \frac{1}{3}$ in Section \ref{sec:brierscorecomp} and $\frac{1}{6}$ in Section \ref{sec:realdata}. The posterior distributions for the GMM follows.

\begin{align}\label{eq:gmm_posterior_app}
    \Pi|\by,\bz &\sim Dirichlet(\alpha_1 + n_1, \ldots \alpha_K + n_K),\\
    \sigma_j^2|\bz,\by &\sim \mathcal{IG}(\hat{\alpha}_j, \hat{\eta}_j),\\
        \mu_j|\bz, \by,\sigma_j &\sim \mathcal{N}(\hat{c}_j, \hat{\sigma}^2_j),\\
    z_i|\by, \mu_j, \sigma_j &\mathop{\sim}^{ind} p(z_i = k | \by, \mu_j, \sigma_j), i = 1,\ldots, n,
\end{align}
where $n_j$ = $\sum_{i = 1}^n \mbox{1}_{z_i = j}$, j = $1, \ldots, n$, $\hat{\alpha}_j = \alpha + (n_j-1)/2$, $\hat{\eta}_j = \eta + \frac{1}{2}\sum_{z_i = j}(y_i - \bar{y_j})^2$, $\hat{c}_j= \frac{n_j\bar{y}_j/\sigma_j^2+c_j/\sigma^2}{n_j/\sigma_j^2+1/\sigma^2}$, $\hat{\sigma}^2_j =\frac{1}{n_j/\sigma_j^2+1/\sigma^2}$. More details of this Gaussian mixture model can be found, for example, in \citet[][Sec.~ 2]{rasmussen1999infinite}.

\cite{stoehr2017}

\bibliographystyle{apalike}
\bibliography{mendeley.bib, additionalrefs.bib}

\end{document}